\documentclass[prd,twocolumn,showpacs,preprintnumbers,superscriptaddress,floatfix,nofootinbibnofootinbib]{revtex4-1}
\usepackage{amssymb}
\usepackage{amsmath,txfonts}
\usepackage{graphicx}
\usepackage{dcolumn}
\usepackage{color}
\usepackage{bm}
\usepackage[subfigure]{graphfig}
\usepackage{makecell}
\usepackage[colorlinks,citecolor=blue]{hyperref}
\pdfpagewidth=\paperwidth
\pdfpageheight=\paperheight
\allowdisplaybreaks
\newcommand*{\slashed}[1]{{#1\!\!\!/}}
\newcommand*{\hc}{\text{H.\,c.}}

\begin{document}

\title{\boldmath The effect of decay cascade via an intermediate resonance in the $\gamma p \to \pi^0 \eta p $ reaction}

\author{Ai-Chao Wang}
\affiliation{College of Science, China University of Petroleum (East China), Qingdao 266580, China}

\author{Neng-Chang Wei}
\email[Corresponding author. Email: ]{weinengchang@htu.edu.cn}
\affiliation{School of Physics, Henan Normal University, Henan 453007, China}

\date{\today}

\begin{abstract}

The $\gamma p \to  \pi^0 \eta p$ reaction has been investigated by the CBELSA/TAPS Collaboration, revealing a narrow structure in the $\eta p$ invariant mass distributions at a mass of $1700$ MeV. In this study, we explore the possibility that the narrow structure is caused by a decay cascade via an intermediate nucleon resonance decaying to $\eta p$ final states. The candidates for the intermediate nucleon resonances are $N(1700)3/2^{-}$ and $N(1710)1/2^{+}$, with masses near the observed structure. We consider the $t$-channel $\rho$- and $\omega$-exchange diagrams, the $u$-channel nucleon-pole exchange diagram, the contact term, and the $s$-channel pole diagrams of nucleon, $\Delta$, and nucleon resonances when constructing the reaction amplitudes to reproduce the stripped individual contribution of the narrow structure. Our analysis indicates that the signature strength of the decay cascade $\gamma p \to \pi^{0}N(1700)3/2^{-} \to \pi^{0}\eta p$ is too weak to reach the experimental curve of the narrow structure due to the small decay branching ratio of $N(1700)3/2^{-}$ to $\eta p$. Although the decay cascade $\gamma p \to \pi^{0}N(1710)1/2^{+} \to \pi^{0}\eta p$ can qualitatively reproduce the experimental curve of the invariant mass distributions, its cross-section width is much larger than that of the corresponding experimental curve. Therefore, we conclude that the decay cascade via an intermediate nucleon resonance could not be the reason leading to the narrow structure in the $\eta p$ invariant mass distributions of the $\gamma p \to  \pi^0 \eta p$ reaction.

\end{abstract}


\maketitle

\section{Introduction}  \label{Sec:intro}

The exploration of structure and spectrum of nucleon resonances ($N^\ast$'s) and $\Delta$ resonances ($\Delta^\ast$'s) has been a focal point in the field of hadron physics. This line of investigation offers valuable insights into the dynamic properties of Quantum Chromodynamics (QCD) in the nonperturbative energy regime. Our current understanding of nucleon resonances primarily stems from studies involving $\pi N$ scattering and $\pi$ photoproduction reactions. However, a substantial number of theoretically predicted nucleon resonances, originating from QCD-inspired phenomenological models \cite{Isgur:1978,Capstick:1986,Loring:2001} and QCD-lattice calculations \cite{Edwards:2013,Engel:2013}, have eluded experimental identification. This discrepancy is particularly evident in the center-of-mass energy (c.m.) 2 GeV region, where the observed nucleon resonances tend to be broader and more susceptible to overlapping. The number of theoretically predicted nucleon resonances surpasses the count of experimentally observed ones, giving rise to the so-called \textit{missing resonance problem} \cite{Koniuk:1980}.

One possible explanation for the lack of experimental detection of missing resonances is their weak couplings to the $\pi N$ channel, making them challenging to observe in such reaction channel. Investigating the $\pi N$-weakly-coupled nucleon resonances through other meson production reactions becomes a viable approach. In recent decades, significant progress has been made in meson photoproductions, both theoretically and experimentally, providing alternative platforms for nucleon resonance studies \cite{hejun:2014,Kim:2014,Zhangy:2021,Wangxiaoyun:2017,Bradford:2006,Mart:2019,
CLAS-beam,Zachariou:2020kkb,Wei:2022,Wang:2022,Anisovich:2017rpe,Wei:2020,Moriya:2013,Wangx:2020,
Zhangyx:2021,Wein:2021,Weinc:2019}. In this work, we employ the Effective Lagrangian method to explore the potential impact of the decay cascade via intermediate nucleon resonance contents in the $\gamma p \to \pi^{0} \eta p$ reaction. As a double-meson emission reaction, the $\gamma p \to \pi^{0} \eta p$ reaction has a higher threshold energy than most single-meson emission reactions, making it more suitable for investigating high-mass nucleon resonances in the less-explored energy region.

Several experimental and theoretical studies have been conducted to investigate the $\gamma p \to \pi^{0}\eta p$ reaction \cite{Doring:2006,Ajaka:2008,Kashevarov:2009,Fix:2013,Doring:2010,Sokhoyan:2020}, providing a foundational understanding of its reaction mechanism. In Ref.~\cite{Doring:2006}, the $\gamma p \to \pi^{0}\eta p$ and $\gamma p \to \pi^{0}K^{0}\Sigma^{+}$ reactions were analyzed using a chiral unitary approach. It was shown that the contribution from the $\Delta(1700)3/2^-$ resonance, followed by its decay into $\Delta(1232)\eta $, plays a dominant role in the $\gamma p \to \pi^{0}\eta p$ reaction, establishing the basic mechanism of this process. The analysis in Ref.~\cite{Ajaka:2008} also emphasized the importance of the $\Delta(1700)3/2^-$ resonance, decaying into $\Delta(1232)\eta$, as a key contributor to the reaction. Ref.~\cite{Kashevarov:2009} presented measurements of total and differential cross sections for the $\gamma p \to \pi^{0}\eta p$ reaction, revealing that in the energy range $E_{\gamma} = 0.95-1.4$ GeV, the reaction is predominantly driven by the excitation and sequential decay of the $\Delta(1700)3/2^-$ resonance. In Ref.~\cite{Fix:2013}, the partial wave structure of the $\gamma p \to \pi^{0}\eta p$ reaction was analyzed over a total center-of-mass energy range from threshold up to $W = 1.9$ GeV. The analysis indicated that the partial wave with quantum numbers $J^P = 3/2^-$ accounts for the largest fraction of the cross section, predominantly saturated by the $\Delta(1700)3/2^-$ resonance. Ref.~\cite{Doring:2010} employed a chiral unitary framework to evaluate polarization observables $I^S$ and $I^C$ for the $\gamma p \to \pi^{0}\eta p$ reaction, further corroborating the significant role of the $\Delta(1700)3/2^-$ resonance. Ref.~\cite{Sokhoyan:2020} reported beam-helicity asymmetry data for the photoproduction of $\pi^0 \eta$ pairs on carbon, aluminum, and lead, demonstrating that the $\gamma p \to \pi^{0}\eta p$ reaction is dominated by the $D_{33}$ partial wave with the $\Delta(1232)\eta $ intermediate state. In summary, these studies consistently agree that the primary mechanism underlying the $\gamma p \to \pi^{0}\eta p$ reaction is dominated by the contribution from the $\Delta(1700)3/2^-$ resonance, decaying via $\Delta(1700) \to \Delta(1232)\eta $. Additionally, several other studies have contributed to the understanding of the $\gamma p \to \pi^{0}\eta p$ reaction \cite{Fix:2022,Martinez:2023,Ishikawa:2021,A2:2018vbv,Debastiani:2017dlz}.

Focusing on the present study, a narrow structure was reported in 2017 in the $\eta N$ invariant mass distribution at $W \sim 1.678$ GeV for the $\gamma N \to \pi \eta N$ reaction, based on data from the GRAAL facility \cite{jiu:2017}. The suggested interpretation was that the observed peak structure corresponds to the nucleon resonance $N(1685)$. Subsequently, in 2021, the CBELSA/TAPS Collaboration conducted a remeasurement of the $\gamma p \to \pi^{0} \eta p$ reaction \cite{xin:2021}. Contrary to the previous findings, they could not confirm the existence of a narrow structure in the $\eta N$ invariant mass distributions at $W \sim 1.678$ GeV. Instead, they observed a narrow structure in the $\eta p$ invariant mass distributions at $W \sim 1.7$ GeV, with a width of $\Gamma\approx35$ MeV, for incident photon energies in the range of $1400-1500$ MeV and a cut of $M_{\pi^{0}p}\leq 1190$ MeV. Furthermore, with increasing incident energy from $1420$ MeV to $1540$ MeV, the structure shifted in mass from $1700$ MeV to $1725$ MeV, and the width increased to about $50$ MeV. The CBELSA/TAPS Collaboration proposed that the most likely explanation for the narrow structure is a triangular singularity in the $\gamma p \to \pi^{0}\eta p$ reaction. Nevertheless, we are curious whether there are other possibilities that could explain the observed structure.

In this study, our objective is to investigate whether the observed narrow structure in the invariant mass distributions of the $\gamma p \to \pi^{0}\eta p$ reaction can be attributed to a decay cascade via an intermediate nucleon resonance decaying into $\eta p$ final states. While the dominant contribution of the decay cascade via the $\Delta(1232)$ resonance in this reaction is well established, we aim to elucidate the role of other potential decay cascades involving nucleon resonances. Considering that the narrow structure is located at $M_{p\eta} = 1700~\text{MeV}/c^2$ and isospin conservation restricts the intermediate resonance to be an isospin-$1/2$ nucleon resonance, we focus on two candidates in close proximity to this mass: $N(1700)3/2^{-}$, a three-star nucleon resonance in the Particle Data Group review (PDG) \cite{PDG:2022}, and $N(1710)1/2^{+}$, a four-star nucleon resonance.

The CBELSA/TAPS Collaboration has extracted the individual contribution of the narrow structure from the $\gamma p \to \pi^{0}\eta p$ reaction \cite{xin:2021}. In this study, we aim to reproduce this extracted narrow structure signal by employing the effective Lagrangian method and considering the decay cascade via possible intermediate resonances. Specifically, to construct the reaction amplitudes for the decay cascade $\gamma p \to \pi^{0} Res.(\eta p) \to \pi^{0}\eta p$, we include contributions from $t$-channel $\rho$- and $\omega$-exchange, $s$-channel nucleon ($N$) and $\Delta$ pole diagrams, $u$-channel nucleon ($N$) pole exchange, and the contact term. For the decay cascade $\gamma p \to \pi^{0} N(1700)3/2^{-} \to \pi^{0}\eta p$, contributions from nucleon resonances $N(1440)1/2^{+}$ and $N(1520)3/2^{-}$ are included. Similarly, for the decay cascade $\gamma p \to \pi^{0} N(1710)1/2^{+} \to \pi^{0}\eta p$, the contribution from the nucleon resonance $N(1535)1/2^{-}$ is considered. It is important to note that our model specifically excludes the dominant background process $\gamma p \to \Delta(1232)\eta \to \pi^{0}\eta p$, as this mechanism is not part of the extracted narrow structure signal.

To focus on the narrow structure, we employ specific operations for the theoretical framework. Following the approach in Ref.~\cite{xin:2021}, an invariant mass distribution cut of $M_{p\pi^{0}} < 1190$ MeV is applied to suppress the dominant decay cascade $\gamma p \to \Delta(1232)\eta \to \pi^{0}\eta p$, ensuring consistency with the experimental data. While the contribution from the $\Delta(1700)3/2^-$ resonance, which decays via $\Delta(1700) \to \Delta(1232)\eta$, is inherently part of the background, it is excluded from our theoretical model as it does not contribute to the narrow structure. Instead, our goal is to reproduce the individual contribution of the narrow structure extracted by the CBELSA/TAPS Collaboration \cite{xin:2021}. As such, the interference effects between the narrow structure and the background terms are beyond the scope of this study. This approach allows us to investigate whether the decay cascade via an intermediate nucleon resonance, $\gamma p \to \pi^{0} Res.\to \pi^{0} \eta p$, can explain the observed narrow structure in the $\eta p$ invariant mass distributions of the $\gamma p \to \pi^{0}\eta p$ reaction, with the results in accordance with the energy-dependent relationship identified by the CBELSA/TAPS Collaboration.

The current paper is organized as follows. In Sec.~\ref{Sec:formalism}, we provide an introduction to the framework of our theoretical model, encompassing the Lagrangians, propagators, form factors, and the reaction amplitudes. In Sec.~\ref{Sec:results}, we show the theoretical results and make a further discussion. Finally, Sec.~\ref{Sec:summary} offers a summary and conclusions.

\section{Formalism}  \label{Sec:formalism}

\begin{figure}[tbp]
\subfigure[~$s$ channel]{
\includegraphics[width=0.45\columnwidth]{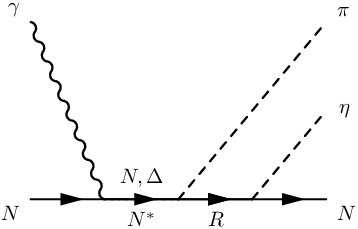}}  {\hglue 0.4cm}
\subfigure[~$t$ channel]{
\includegraphics[width=0.45\columnwidth]{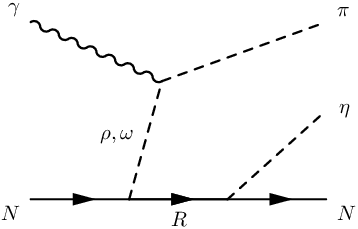}} \\[6pt]
\subfigure[~$u$ channel]{
\includegraphics[width=0.45\columnwidth]{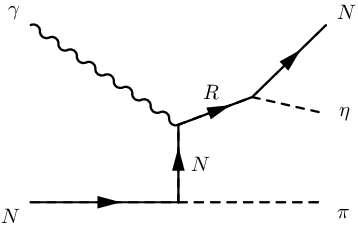}} {\hglue 0.4cm}
\subfigure[~contact term]{
\includegraphics[width=0.45\columnwidth]{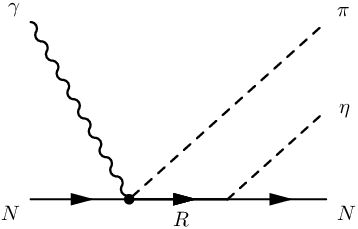}}
\caption{Generic structure of the amplitudes for the decay cascade $\gamma N\to \pi Res. \to \pi \eta N$ reaction. For the case of $\gamma p \to \pi^{0} N(1700)3/2^{-} \to \pi^{0}\eta p$ reaction, the $R$ symbol represents the intermediate resonance $N(1700)3/2^-$, and the $N^{*}$ represents the $N(1440)1/2^{+}$ and $N(1520)3/2^{-}$ resonances. For the case of $\gamma p \to \pi^{0} N(1710)1/2^{+} \to \pi^{0}\eta p$ reaction, the $R$ symbol represents the intermediate resonance $N(1710)1/2^{+}$, and the $N^\ast$ represents the $N(1535)1/2^{-}$ resonance. Time proceeds from left to right. }
\label{FIG:feymans}
\end{figure}

The full reaction amplitude for the decay cascade reaction $\gamma p \to \pi^{0} Res. \to \pi^{0} \eta p$ can be expressed in the following form:
\begin{equation}
M = \bar{u}(p_{5},\lambda_{p'})A_{\mu}M^{\mu}u(p_{2},\lambda_{p}), \label{eq:amplitude}
\end{equation}
where $\lambda_{p}$ and $\lambda_{p'}$ are the helicities of the incoming proton and outgoing proton, respectively; $\mu$ is the Lorentz index of the incoming photon, and $M^{\mu}$ is the full amplitude with outer lines omitted. The contributions considered in constructing the reaction amplitudes include: i) $s$-channel $N$, $\Delta$, and $N^\ast$ pole diagrams, ii) $t$-channel $\rho$- and $\omega$-exchange diagrams, iii) $u$-channel $N$-pole exchange diagram, and iv) contact term diagram. In the decay cascade $\gamma p \to \pi^{0} N(1700)3/2^{-} \to \pi^{0}\eta p$, $N^\ast$ denotes the nucleon resonances $N(1440)1/2^{+}$ and $N(1520)3/2^{-}$, one of which, together with $\pi$, constitutes the decay mode of $N(1700)3/2^{-}$ as presented in the PDG review \cite{PDG:2022}. Similarly, for the decay cascade $\gamma p \to \pi^{0} N(1710)1/2^{+} \to \pi^{0}\eta p$, $N^\ast$ represents the nucleon resonance $N(1535)1/2^{-}$.

In the remainder of this section, we present the effective Lagrangians, propagators, and phenomenological form factors used in this work to construct the reaction amplitudes. Additionally, we outline the final forms of the reaction amplitudes for all the particle-exchange cases in the two cascade decay reactions.

\subsection{Effective Lagrangians} \label{Sec:Lagrangians}

The effective Lagrangians utilized in our present work are provided below. For convenience, we define the field-strength tensor for the electromagnetic field and the vector-meson field as
\begin{eqnarray}
F^{\mu\nu} &=& \partial^{\mu}A^\nu-\partial^{\nu}A^\mu,\nonumber \\[6pt]
V^{\mu\nu} &=& \partial^{\mu}V^\nu-\partial^{\nu}V^\mu,
\end{eqnarray}
where $V$ represents the $\rho$ or $\omega$ vector meson.

\subsubsection{The effective Lagrangians for electromagnetic vertexes in the $\gamma N \to \pi N(1700)3/2^- \to \pi \eta N$ process}

\begin{eqnarray}
{\cal L}_{ NN \gamma} &=& -\,e \bar{N} \left[ \left( \hat{e} \gamma^\mu - \frac{ \hat{\kappa}_N} {2M_N}\sigma^{\mu \nu}\partial_\nu\right) A_\mu\right] N, \label{Lags1}\\[6pt]
{\cal L}_{\pi\rho\gamma } &=&e \frac{g_{\pi\rho\gamma}}{M_{\pi}}\varepsilon^{ \alpha \mu \lambda \nu}(\partial_{\alpha }A_{\mu})(\partial_{\lambda}\pi)\rho_{\nu },\label{Lags2}\\[6pt]
{\cal L}_{\pi\omega \gamma} &=&e \frac{g_{\pi\omega\gamma}}{M_{\pi}}\varepsilon^{ \alpha \mu \lambda \nu}(\partial_{\alpha }A_{\mu})(\partial_{\lambda}\pi^{0})\omega_{\nu },\label{Lags3}\\[6pt]
{\cal L}_{\Delta N \gamma} &=& - ie\frac{g^{(1)}_{\Delta N \gamma}}{2M_N}\bar{\Delta}_{\mu}\gamma_\nu \gamma_5 F^{\mu \nu} N \nonumber \\
&& +\,e\frac{g^{(2)}_{\Delta N \gamma}}{\left(2M_N\right)^2} \bar{\Delta}{_\mu} \gamma_5 F^{\mu \nu}\partial_\nu N + \hc,\label{Lags4}\\[6pt]
{\cal L}_{RN\gamma} &=& -\, ie\frac{g_{RN\gamma}^{(1)}}{2M_N}\bar{R}_\mu \gamma_\nu F^{\mu\nu}N \nonumber \\
&&+\, e\frac{g_{RN\gamma}^{(2)}}{\left(2M_N\right)^2}\bar{R}_\mu F^{\mu \nu}\partial_\nu N + \hc,\\[6pt]
\mathcal{L}_{R N\pi\gamma} &=&  -\, i e \frac{ g_{R N\pi}}{M_\pi} \bar{R}^{\mu} \gamma_{5}A_{\mu}\pi N  + \hc,   \\[6pt]
{\cal L}_{ N^{*}N \gamma}^{1/2^{+}} &=& e \frac{ g_{ N^{*} N \gamma}^{(1)}}{2M_N}\bar{N}^{*} \sigma^{\mu \nu}(\partial_\nu A_\mu) N+ \hc,\label{Lags5}\\[6pt]
{\cal L}_{N^{*}N\gamma}^{3/2^{-}} &=& -\, ie\frac{g_{N^{*}N\gamma}^{(1)}}{2M_N}\bar{N}^{*}_\mu \gamma_\nu  F^{\mu\nu}N \nonumber \\
&& +\, e\frac{g_{N^{*}N\gamma}^{(2)}}{\left(2M_N\right)^2}\bar{N}^{*}_\mu  F^{\mu \nu}\partial_\nu N + \hc,
\end{eqnarray}
where $e$ is the elementary charge unit; $\hat{e}$ and $R$ represent the charge operator and nucleon resonance $N(1700)3/2^{-}$, respectively; The anomalous magnetic moments are defined as $\hat{\kappa}_N = \kappa_p(1+\tau_3)/2 + \kappa_n(1-\tau_3)/2$, with the anomalous magnetic moments $\kappa_p=1.793$ and $\kappa_n=-1.913$. $M_N$ stands for the mass of $N$. The coupling constant $g_{\pi \rho\gamma}$  and $g_{\pi \omega\gamma}$ are determined by caculating the vector meson radiative decay width
\begin{eqnarray}
\Gamma _{V \rightarrow \pi \gamma }=\frac{e^2}{4\pi}\frac{g_{V \pi\gamma }^{2}}{24M_{V}^{3}M_{\pi}^2}(M_{V}^{2}-M_{\pi}^{2})^{3},
\end{eqnarray}
where $V$ stands for the vector meson $\rho$ or $\omega$. Utilizing the decay width values $\Gamma _{\rho^{0}\rightarrow \pi^{0} \gamma }\simeq0.070$ MeV and $\Gamma _{\omega\rightarrow \pi^{0} \gamma } \simeq0.72$ MeV from the PDG review \cite{PDG:2022}, we obtain $g_{\pi \rho\gamma}=0.099$ and $g_{\pi \omega\gamma}=0.31$. The coupling constants for $RN\gamma$ are determined by PDG values of $N(1700)3/2^- \to N\gamma$ helicity amplitudes: $A_{1/2}=0.032 ~{\rm GeV^{-1/2}}$ and $A_{3/2}=0.034~{\rm GeV^{-1/2}}$, yielding $g^{(1)}_{R N \gamma}=0.405$ and $g^{(2)}_{R N \gamma}=-0.986$. Similarly, $\Delta N\gamma$ couplings are determined by PDG values of $\Delta(1232)3/2^+ \to N\gamma$ helicity amplitudes: $A_{1/2}=-0.135~{\rm GeV^{-1/2}}$ and $A_{3/2}=-0.255~{\rm GeV^{-1/2}}$, leading to $g^{(1)}_{\Delta N \gamma}=-4.17$ and $g^{(2)}_{\Delta N \gamma}=4.32$. The $s$-channel $N(1440)1/2^+$ and $N(1520)3/2^-$ pole diagrams are considered due to their coupling with $\pi N(1700)3/2^{-}$ as stated in the PDG review \cite{PDG:2022}. The coupling constant of $N(1440)1/2^{+} N \gamma$ vertex is taken as $g_{ N^{*} N \gamma}^{(1)}=0.505$, obtained from the helicity amplitude $A_{1/2}=-0.065 ~{\rm GeV^{-1/2}}$. The coupling constant of $N(1520)3/2^{-} N \gamma$ vertex is taken as $g_{ N^{*} N \gamma}^{(1)}=-4.86$ and $g_{ N^{*} N \gamma}^{(2)}=5.27$, derived from the helicity amplitudes $A_{1/2}=-0.025 ~{\rm GeV^{-1/2}}$ and $A_{3/2}=0.14 ~{\rm GeV^{-1/2}}$ \cite{PDG:2022}.

It is important to note that in the present work, the couplings of $N^* N \gamma$ are determined from the helicity amplitudes using the following relations, with all helicity amplitudes taken from the PDG review \cite{PDG:2022}.

For a spin-1/2 resonance with parity $P=\pm 1$, the relation is given by
\begin{eqnarray}
\bar{g}_{1}=-\frac{1}{e}\sqrt{\frac{M_{N}}{kM_{N^*}}}A_{1/2},
\end{eqnarray}
where
\begin{equation}
\bar{g}_{1}=\frac{g^{(1)}_{N^*Nr}}{2M_{N}}.
\end{equation}

For a spin-3/2 resonance with parity $P=\pm 1$, the relations are
\begin{eqnarray}
\bar{g}_{1}&=&\frac{1}{e}\frac{1}{\sqrt{2}k}\sqrt{\frac{M_{N}}{k M_{N^*}}}\left[\sqrt{3}A_{1/2}\pm A_{3/2}\right](M_{N^*}\mp M_{N}),\\[6pt]
\bar{g}_{2}&=&-\frac{1}{e}\frac{\sqrt{2}}{k M_{N^*}}\sqrt{\frac{M_{N}}{k M_{N^*}}}\left[\sqrt{3}M_{N^*}A_{1/2}- M_{N}A_{3/2}\right],
\end{eqnarray}
where
\begin{eqnarray}
\bar{g}_{1}=\frac{g^{(1)}_{N^*Nr}}{2M_{N}},\qquad \bar{g}_{2}=\frac{g^{(2)}_{N^*Nr}}{(2M_{N})^2}.
\end{eqnarray}

\subsubsection{The effective Lagrangians for electromagnetic vertexes in the $\gamma N \to \pi N(1710)1/2^+ \to \pi \eta N$ process}

The Lagrangians outlined in Eq.~(\ref{Lags1}-\ref{Lags4}) are also needed for constructing the amplitudes in the $\gamma N \to \pi N(1710)1/2^+$ reaction, and we refrain from repeating them here. Additionally, the other required Lagrangians are presented below
\begin{eqnarray}
{\cal L}_{ RN \gamma} &=& e \frac{ g_{ R N \gamma}^{(1)}}{2M_N}\bar{R}\sigma^{\mu \nu}(\partial_\nu A_\mu) N+ \hc, \\[6pt]
\mathcal{L}_{R N\pi\gamma} &=& i e \frac{g_{R N\pi}}{M_N+M_R} \bar{R} \gamma_{5}\gamma^{\mu}A_{\mu}\pi N  + \hc,   \\[6pt]
{\cal L}_{ N^{*}N \gamma}^{1/2^{-}} &=& e \frac{ g_{ N^{*} N \gamma}^{(1)}}{2M_N}\bar{N}^{*}\gamma_{5}\sigma^{\mu \nu}(\partial_\nu A_\mu) N+ \hc,
\end{eqnarray}
where $R$ represents the nucleon resonance $N(1710)1/2^{+}$. Additionally, the $s$-channel $N(1535)1/2^-$ pole diagram is considered due to their coupling with $\pi N(1710)1/2^{+}$ as stated in the PDG review \cite{PDG:2022}. The coupling constants for $N(1710)1/2^{+} N\gamma$ and $N(1535)1/2^- N\gamma$ vertices are estimated as $g_{ R N \gamma}^{(1)}=0.134$ and $g_{ N^* N \gamma}^{(1)}=0.619$ based on their decay branching widths ${\Gamma}_{N(1710)1/2^{+} \rightarrow N\gamma}\simeq80\times0.04\%=0.032$ MeV and ${\Gamma}_{N(1535)1/2^{+} \rightarrow N\gamma}\simeq150\times0.23\%=0.345$ MeV.

To reconcile the observed narrow structure width of approximately $35-50$ MeV with the relatively broader PDG-reported widths of the candidate resonances $N(1700)3/2^-$ and $N(1710)1/2^{+}$, we assume a total width of $80$ MeV for both resonances in our analysis. The PDG-reported total widths, ranging from $100$ to $300$ MeV for $N(1700)3/2^-$ and from $80$ to $200$ MeV for $N(1710)1/2^{+}$, are derived from a comprehensive set of photo- and hadroproduction data. Whether the assumed values in our study are consistent with the full dataset from which the PDG widths were determined remains an open question. It should be emphasized that the observed narrow structure in the $\eta p$ invariant mass spectrum, with a width of approximately $35-50$ MeV, arises from the interference effects among the relevant reaction amplitudes rather than directly reflecting the intrinsic widths of the contributing resonances.

The width formula for the nucleon resonance with spin $1/2$ decaying into $N\gamma$ is denoted as
\begin{eqnarray}
\Gamma _{N^*\rightarrow N \gamma }^{1/2\pm}&=&\frac{e^2}{4\pi}\frac{g_{ N^* N \gamma}^{(1)^2}}{(2M_N)^2}4k^3,
\end{eqnarray}
which is applicable to both nucleon resonances $N(1710)1/2^{+}$ and $N(1535)1/2^-$.

\subsubsection{The effective Lagrangians for meson-baryon vertexes in the $\gamma N \to \pi N(1700)3/2^- \to\pi \eta N$ process}

\begin{eqnarray}
\mathcal{L}_{R N\pi} &=& \frac{g_{R N\pi}}{M_\pi} \bar{R}^{\mu} \gamma_{5}\left( \partial_\mu {\pi} \right) N  + \hc,   \\[6pt]
\mathcal{L}_{R N\eta} &=& -\frac{g_{R N\eta}}{M_\eta} \bar{N} \gamma_{5}\left( \partial_\mu {\eta} \right) R^{\mu}  + \hc,    \\[6pt]
\mathcal{L}_{R\Delta \pi} &=& \frac{g_{R\Delta \pi}^{(1)}}{M_\pi} \bar{\Delta}_\mu  \gamma^\alpha \left(\partial_\alpha \pi\right) R^\mu \nonumber \\
&& + \, i \frac{g_{R\Delta \pi}^{(2)}}{M_{\pi}^2} \bar{\Delta}_\alpha  \left(\partial^\mu \partial^\alpha \pi\right)  R_\mu + \hc,   \\[6pt]
{\cal L}_{R N{\rho}} &=& -\, i\frac{g_{R N \rho}^{(1)}}{2M_{N}}{\bar R}_\mu \gamma_\nu  {\rho}^{\mu \nu}N   +\, \frac{g_{R N \rho}^{(2)}}{\left(2M_{N}\right)^2}{\bar R}_\mu  {\rho}^{\mu \nu}\partial_\nu N \nonumber \\
&& +\, \frac{g_{R N \rho}^{(3)}}{\left(2M_{N}\right)^2}{\bar R}_\mu \left(\partial_\nu {\rho^{\mu \nu}}\right) N + \hc, \\[6pt]
{\cal L}_{R N{\omega}} &=& -\, i\frac{g_{R N \omega}^{(1)}}{2M_{N}}{\bar R}_\mu \gamma_\nu  {\omega}^{\mu \nu}N   +\, \frac{g_{R N \omega}^{(2)}}{\left(2M_{N}\right)^2}{\bar R}_\mu  {\omega}^{\mu \nu}\partial_\nu N \nonumber \\
&& +\, \frac{g_{R N \omega}^{(3)}}{\left(2M_{N}\right)^2}{\bar R}_\mu \left(\partial_\nu {\omega^{\mu \nu}}\right) N + \hc, \\[6pt]
\mathcal{L}_{N N\pi} &=& - g_{NN\pi}\bar{N}  \gamma_{5} \left[\left( i\lambda +\frac{1-\lambda}{2M_N}\slashed{\partial}\right)\pi \right]N, \\[6pt]
\mathcal{L}_{N^{*} R\pi}^{1/2^{+}} &=& \frac{g_{N^{*} R\pi}^{(1)}}{M_{\pi}}\bar{R}^{\mu}
\gamma_{5}(\partial_{\mu}\pi)N^{*} + \hc, \\[6pt]
\mathcal{L}_{N^{*} R\pi}^{3/2^-} &=& \frac{g_{N^{*} R\pi}^{(1)}}{M_{\pi}} \bar{R}^{\mu}  \gamma^\alpha \gamma_{5} \left(\partial_\alpha \pi\right) N^{*}_\mu \nonumber \\
&& + \, i \frac{g_{N^{*} R\pi}^{(2)}}{M_{\pi}^2} \bar{R}_{\alpha}  \gamma_{5} \left(\partial^\mu \partial^\alpha \pi\right)  R_\mu + \hc,
\end{eqnarray}
where the symbol $R$ represents the nucleon resonance $N(1700)3/2^{-}$. The coupling constants $g_{R N\pi}$, $g_{R N\eta}$ and $g_{R \Delta\pi}^{(1)}$ can be determined by the branching width of $N(1700)3/2^{-}$ decaying to $N\pi$, $N\eta$ and $\Delta\pi$, respectively. The width formulas are denoted as follows
\begin{eqnarray}
\Gamma _{R\rightarrow N P }&=&\frac{g_{RNP}^2}{4\pi}\frac{1}{3}\frac{k^{3}}{M_{R}M_{P}^{2}}(E_{N}-M_{N}),\\[6pt]
\Gamma_{R\rightarrow \Delta\pi} &=& \frac{1}{36\pi}
\frac{k}{M_R M_{\Delta}^2} (E_{\Delta} + M_{\Delta})\nonumber \\
&&\times\Biggl\{ \frac{g_{R\Delta \pi}^{(1)^2}}{M_{\pi}^2} (M_R - M_{\Delta})^2 \nonumber \\
&&\times\left( 2
E_{\Delta}^2 + 2 E_{\Delta} M_{\Delta} + 5 M_{\Delta}^2
\right) \nonumber \nonumber \\
&&+ 2 \frac{g_{R\Delta \pi}^{(1)}g_{R\Delta \pi}^{(2)}}{M_{\pi}^3} M_R k^2 (M_R - M_{\Delta}) (2E_{\Delta} +
M_{\Delta}) \nonumber \\
&&+ 2 \frac{g_{R\Delta \pi}^{(2)^2}}{M_{\pi}^4} M_R^2 k^4 \Biggr\},
\end{eqnarray}
where $P$ represents the pseudoscalar meson $\pi$ or $\eta$. Given the assumption in the present article that the total decay width of $N(1700)3/2^{-}$ is $80$ MeV, the coupling constants for the $RN\pi$, $RN\eta$, and $R\Delta\pi$ vertices are estimated as $g_{RN\pi}=0.588$, $g_{RN\eta}=2.08$, and $g_{R\Delta\pi}^{(1)}=-0.386$ based on the decay branching widths $\Gamma _{R\rightarrow N \pi }\simeq80\times12\%=9.6$ MeV, $\Gamma _{R\rightarrow N \eta}\simeq80\times1.5\%=1.2$ MeV, and $\Gamma _{R\rightarrow \Delta \pi }\simeq80\times70\%=56$ MeV. These branching ratios are referenced from the PDG review \cite{PDG:2022}. For simplicity, we omit the $g^{(2)}$ and $g^{(3)}$ terms in the Lagrangians of $R\Delta\pi$, $R N \rho$, $R N \omega$, and $N^{*}R\pi$ couplings due to the lack of experimental information. Consequently, the energy-dependent width formula for $RNV$ coupling can be expressed as
\begin{eqnarray}
\Gamma _{R\rightarrow NV }(\sqrt{s})&&=\frac{1}{12\pi}\frac{k}{M_R}(E_N+M_N)
\nonumber \\
&&\times\frac{g_{RNV}^{(1)^2}}{4M_N^2}\left[2E_N(E_N-M_N)+(M_R-M_N)^2+2M_V^2\right],\nonumber \\
\end{eqnarray}
where $V$ represents the vector meson $\rho$ or $\omega$. Additionally, for the decay process $R\rightarrow \rho N/ \omega N$, considering the mass of the nucleon resonance $N(1700)3/2^-$ lying close to the $\rho N/ \omega N$ threshold, it is necessary to account for the effect of finite decay width. Referring to previous studies~\cite{kuandu1,kuandu2,kuandu3}, the final formula for the decay width of $RNV$ is given by
\begin{eqnarray}
\Gamma _{R\rightarrow NV }&=&-\frac{1}{\pi}\int_{(M_R-2\Gamma_R)^2}^{(M_R+2\Gamma_R)^2}\Gamma _{R\rightarrow NV }(\sqrt{s})\nonumber \\
&&\times\Theta(\sqrt{s}-M_N-M_V)\nonumber \\
&&\times \rm{ Im}\it\left\{\frac{{\rm 1}}{s-M_{R}^2+iM_{R}\Gamma_{R}}\right\} ds.
\label{eq:width_function}
\end{eqnarray}

The coupling constants $g_{R N \rho}=5.57$ and $g_{R N \omega}=4.89$ are determined using the formulas above, along with the estimated branching widths $\Gamma _{R\rightarrow N \rho}\simeq80\times38\%=30.4$ MeV and $\Gamma _{R\rightarrow N \omega}\simeq80\times22\%=17.6$ MeV. It is important to note that all the decay branching ratios in this work adopt the middle values within the suggested intervals by the PDG review \cite{PDG:2022}. Following Ref.~\cite{ff3}, a pure pseudovector coupling ($\lambda=0$) is assumed for the $NN\pi$ vertex, with the coupling constant $g_{NN\pi}=0.989$ adopted from Ref.~\cite{Ronchen:2013}, determined by the SU(3) flavor symmetry. Additionally, the coupling constants for $N(1440)1/2^+ R\pi$ and $N(1520)3/2^{-} R\pi$ are considered. The coupling constant $g_{N^{*}R\pi}^{(1)}$ for $N(1440)1/2^+ R\pi$ is taken as $7.23$, leading from the decay branching width ${\Gamma}_{R\rightarrow N(1440)1/2^{+} \pi}\simeq80\times7\%=5.6$ MeV. Similarly, for $N(1520)3/2^{-} R\pi$, the coupling constant $g_{ N^{*}R\pi}^{(1)}$ is calculated as $-0.759$, derived from the decay branching width ${\Gamma}_{ R\rightarrow N(1535)1/2^-\pi}\simeq80\times4\%=3.2$ MeV. The width formulas are denoted as follows

\begin{eqnarray}
\Gamma _{R\rightarrow N(1440)1/2^{+} \pi }&=&\frac{g_{N^{*}R\pi}^{(1)^{2}}}{4\pi}\frac{1}{3}\frac{k^{3}}{M_{R}M_{\pi}^{2}}(E_{N^*}-M_{N^*}),\\[6pt]
\Gamma_{R\rightarrow N(1520)3/2^{-}\pi} &=& \frac{1}{36\pi}
\frac{k}{M_R M_{N^{*}}^2} (E_{N^{*}} - M_{N^{*}})\nonumber \\
&&\times\Biggl\{ \frac{g_{N^{*}R\pi}^{(1)^2}}{M_{\pi}^2} (M_R + M_{N^{*}})^2 \nonumber \\
&&\times\left( 2
E_{N^{*}}^2 - 2 E_{N^{*}} M_{N^{*}} + 5 M_{N^{*}}^2
\right) \nonumber \nonumber \\
&&+ 2 \frac{g_{N^{*}R\pi}^{(1)}g_{N^{*}R\pi}^{(2)}}{M_{\pi}^3} M_R k^2 (M_R + M_{N^{*}}) (2E_{N^{*}} -M_{N^{*}}) \nonumber \\
&&+ 2 \frac{g_{N^{*}R\pi}^{(2)^2}}{M_{\pi}^4} M_R^2 k^4 \Biggr\}.
\end{eqnarray}

\subsubsection{The effective Lagrangians for meson-baryon vertexes in the $\gamma N \to \pi N(1710)1/2^+  \to\pi \eta N$ process}

\begin{eqnarray}
\mathcal{L}_{R N\pi} &=& - \,g_{R N\pi} \bar{R} \gamma_{5}\left[\left(i\lambda +\frac{1-\lambda}{M_N+M_R}\slashed{\partial}\right)\pi\right] N  + \hc,   \\[6pt]
\mathcal{L}_{R N\eta} &=& - \,g_{R N\eta} \bar{R} \gamma_{5}\left[\left(i\lambda +\frac{1-\lambda}{M_N+M_R}\slashed{\partial}\right)\eta\right] N  + \hc,   \\[6pt]
\mathcal{L}_{R\Delta \pi} &=& \frac{g_{R\Delta \pi}^{(1)}}{M_\pi} \bar{R}\left(\partial^\nu \pi\right) {\Delta}_\nu + \hc, \\[6pt]
{\cal L}_{R N{\rho}} &=&-\, \frac{g_{R N \rho}}{2M_{N}}{\bar R} \left[\left(\frac{\gamma_{\mu}\partial^{2}}{M_R-M_N}
+i\partial_{\mu}\right)V^{\mu}\right]N +  \hc,   \\[6pt]
{\cal L}_{R N{\omega}} &=&-\, \frac{g_{R N \omega}}{2M_{N}}{\bar R} \left[\left(\frac{\gamma_{\mu}\partial^{2}}{M_R-M_N}
+i\partial_{\mu}\right)V^{\mu}\right]N +  \hc,   \\[6pt]
\mathcal{L}_{N^{*} R\pi}^{1/2^-} &=& - \, g_{N^{*} R\pi}^{(1)}\bar{R} \left[\left( i\lambda +\frac{1-\lambda}{M_{N^{*}}-M_R}\slashed{\partial}\right)\pi \right]N^{*}+ \hc,
\end{eqnarray}
where the symbol $R$ denotes the nucleon resonance $N(1710)1/2^{+}$. Pure pseudovector-type couplings are employed for the $RN\pi$ and $RN\eta$ vertices. The coupling constants $g_{R N\pi}$, $g_{R N\eta}$, and $g_{R \Delta\pi}^{(1)}$ are determined through the branching width of $N(1710)1/2^{+}$ decaying into $N\pi$, $N\eta$, and $\Delta\pi$, respectively. The width formulas are expressed as follows
\begin{eqnarray}
\Gamma _{R\rightarrow N P }&=&\frac{g_{RNP}^2}{4\pi}\frac{k}{M_{R}}(E_{N}-M_{N}),\\[6pt]
\Gamma_{R\rightarrow \Delta\pi} &=& \frac{g_{R\Delta \pi}^{(1)^2}}{6\pi}
\frac{k^3 M_{R}}{M_{\pi}^2 M_{\Delta}^2} (E_{\Delta} + M_{\Delta}),
\end{eqnarray}
where $P$ denotes the pseudoscalar meson $\pi$ or $\eta$. The coupling constants for $RN\pi$, $RN\eta$, and $R\Delta\pi$ vertices are specified as $g_{RN\pi}=1.47$, $g_{RN\eta}=3.81$, and $g_{R\Delta\pi}^{(1)}=0.0973$ based on the decay branching widths $\Gamma _{R\rightarrow N \pi }\simeq80\times12.5\%=10$ MeV, $\Gamma _{R\rightarrow N \eta}\simeq80\times30\%=24$ MeV, and $\Gamma _{R\rightarrow \Delta \pi }\simeq80\times6\%=4.8$ MeV. The reaction energy-dependent width formula for the $RNV$ coupling is simplified to
\begin{eqnarray}
\Gamma _{R\rightarrow NV }(\sqrt{s})&&=\frac{g_{RNV}^{(1)^2}}{16\pi}\frac{k(E_N-M_N)}{M_R M_N^2}
\nonumber \\
&&\times\left\{\frac{M_V^2}{(M_R-M_N)^2}\left[(M_R+M_N)^2+2M_V^2\right]\right\},
\end{eqnarray}
where $V$ represents the vector meson $\rho$ or $\omega$. It is essential to consider the effect of finite decaying width \cite{kuandu1,kuandu2,kuandu3}, and the decay width formula remains consistent with Eq.~(\ref{eq:width_function}). Consequently, the coupling constants $g_{R N \rho}=8.47$ and $g_{R N \omega}=4.16$ are obtained from the estimated branching widths $\Gamma _{R\rightarrow N \rho}\simeq80\times17\%=13.6$ MeV and $\Gamma _{R\rightarrow N \omega}\simeq80\times3\%=2.4$ MeV. A pure pseudovector-type coupling is also applied to the $N(1440)1/2^+ R\pi$ vertex. The coupling constant for the $N(1535)1/2^- R\pi$ vertex is set as $g_{N^{*}R \pi}^{(1)}=0.864$, derived from the decay branching width ${\Gamma}_{R\rightarrow N(1535)1/2^{-} \pi}\simeq80\times15\%=12$ MeV. The width formula is expressed as
\begin{eqnarray}
\Gamma _{R\rightarrow N(1535)1/2^{-} \pi }&=&\frac{g_{N^{*}R\pi}^{(1)^2}}{4\pi}\frac{k}{M_{R}}(E_{N}+M_{N}).
\end{eqnarray}

\subsection{Propagators} \label{prop}

For the $s$- and $u$-channel $N$, the propagator is represented as
\begin{eqnarray}
S_{1/2}(p) &=&  \frac{i}{\slashed{p}-M_{N}}.
\end{eqnarray}

For the $s$-channel spin-1/2 resonance $N(1440)1/2^+$ and $N(1535)1/2^-$, the propagator is denoted as
\begin{eqnarray}
S_{1/2}(p) &=&  \frac{i}{\slashed{p}-M_{R}+i\Gamma_{R}/2},
\end{eqnarray}
where $M_{R}$ and $\Gamma_{R}$ represent the mass and width of resonance, respectively.

For the $t$-channel $\rho$ or $\omega$ vector meson exchange, the propagator is denoted as
\begin{eqnarray}
S_{1}(p) &=&  \frac{i(-g^{\mu\nu}+p^{\mu}p^{\nu}/M_V^2)}{p^2-M_V^2}.
\end{eqnarray}

For the $s$-channel $\Delta$ contribution and subsequent decay process of the nucleon resonance $N(1700)3/2^-$, the propagator for a particle with spin $3/2$ is given by
\begin{eqnarray}
S_{3/2}(p) &=&  \frac{i}{\slashed{p} - M_R + i \Gamma_{R}/2} \left( \tilde{g}_{\mu \nu} + \frac{1}{3} \tilde{\gamma}_\mu \tilde{\gamma}_\nu \right),
\end{eqnarray}
where
\begin{eqnarray}
\tilde{g}_{\mu \nu} &=& -\, g_{\mu \nu} + \frac{p_{\mu} p_{\nu}}{M_R^2}, \\[6pt]
\tilde{\gamma}_{\mu} &=& \gamma^{\nu} \tilde{g}_{\nu \mu} = -\gamma_{\mu} + \frac{p_{\mu}\slashed{p}}{M_R^2}.
\end{eqnarray}

\subsection{Form Factors} \label{ff}

To parametrize the structure of the hadrons and normalize the behavior of the production amplitude, a form factor needs to be attached to each hadronic vertex. Although two hadronic vertices exist in each reaction amplitude, only one form factor of the following type is introduced here. Following Refs.~\cite{ff3,ac2018}, for intermediate baryon exchange, the form factor is taken as
\begin{equation}
f_{s(u)}\left(p^2\right) = \left( \frac{\Lambda_B^4}{\Lambda_B^4+\left(p^2-M_B^2\right)^2} \right)^2,
\end{equation}
where $p$, $M_B$, and $\Lambda_B$ denote the four-momentum, the mass, and the cutoff mass of the exchanged baryon $B$, respectively. The same cutoff mass is employed for the $s$- and $u$-channel $N$. The particle mass in the form factor of the contact term is set to the same value as that of $N$. For intermediate meson exchange, the form factor is taken as
\begin{equation}
f_t\left(q^2\right) =  \left(\frac{\Lambda_M^2-M_M^2}{\Lambda_M^2-q^2}\right)^2,
\end{equation}
where $q$, $M_M$, and $\Lambda_M$ represent the four-momentum, the mass, and the cutoff mass for the exchanged meson $M$, respectively. The same cutoff mass is taken for the $t$-channel $\rho$ and $\omega$ exchanges.

\subsection{Reaction amplitudes}

With all the Lagrangians, propagators, and form factors introduced above, the standard amplitudes for the decay cascade $\gamma p \to \pi^{0} N(1700)3/2^{-} \to \pi^{0}\eta p$ can be denoted as
\begin{eqnarray}
M^{\mu}=T_{\nu}M^{\mu\nu},
\end{eqnarray}
where $T_{\nu}$ represents the $N(1700)3/2^{-} \to \eta p$ decaying part. This part is the same for all individual amplitudes of $t$-, $u$-, $s$-channel, and contact term cases, and it is specifically represented as
\begin{eqnarray}
T_{\nu}=\frac{g_{RN\eta}}{m_{\eta}} \gamma_{5}p_{4}^{\beta}S^{3/2}_{\beta\nu}(p_{R}).
\end{eqnarray}
And $M^{\mu\nu}$ represents the remaining parts of the leg-amputated amplitudes $M^{\mu}$, which are represented as follows
\begin{eqnarray}
M_{N(s)}^{\mu\nu}&=& i e \frac{g_{RN\pi}}{m_{\pi}}\gamma_{5}p_{3}^{\nu}S^{1/2}(p_{s}) \nonumber \\
&&\times \, \left[\gamma^{\mu}+\frac{i\kappa_{N}}{2m_{N}}\sigma^{\mu\alpha}p_{1\alpha}\right]f_{s}(p_s,m_N),,   \\[6pt]
M_{N(u)}^{\mu\nu}&=& -\,i \frac{g_{NN\pi}}{2m_N}(p_{1}^{\nu}g^{\mu\alpha}-p_{1}^{\alpha}g^{\mu\nu})\left[\frac{eg^{(1)}_{RN\gamma}}{2m_{N}}
\gamma_{\alpha}+\frac{eg^{(2)}_{RN\gamma}}{(2m_{N})^2}p_{u\alpha} \right] \nonumber \\
&& \times \, S^{1/2}(p_{u})\gamma_{5}\slashed{p}_{3}f_{u}(p_u,m_N),  \\[6pt]
M_{\Delta}^{\mu\nu}&=& i\frac{g_{R\Delta\pi}}{m_\pi}\slashed{p}_{3}g^{\lambda\nu}S^{3/2}_{\lambda\delta}(p_s)
(p_{1}^{\delta}g^{\mu\beta}-p_{1}^{\beta}g^{\mu\delta})\nonumber \\
&&\times \,\left[\frac{eg^{(1)}_{\Delta N\gamma}}{2m_{N}}
\gamma_{\beta}\gamma_{5}+\frac{eg^{(2)}_{\Delta N\gamma}}{(2m_{N})^2}\gamma_{5}p_{2\beta} \right] f_{s}(p_s,m_\Delta),  \\[6pt]
M_{N(1440)}^{\mu\nu}&=& -\, e \frac{g_{N^{*}R\pi}^{(1)}g_{N^{*}N\gamma}^{(1)}}{2m_{\pi}m_N}\gamma_{5}p_3^{\nu}S^{1/2}(p_s)\nonumber \\
&&\times \, \sigma^{\mu\alpha}p_{1\alpha}f_s(p_{s},m_{N(1440)}), \\[6pt]
M_{N(1520)}^{\mu\nu}&=& i\frac{g_{N^{*}R\pi}^{(1)}}{m_\pi}\slashed{p}_{3}\gamma_{5}g^{\lambda\nu}S^{3/2}
_{\lambda \delta}(p_s)(p_{1}^{\delta}g^{\mu\beta}-p_1^{\beta}g^{\mu\delta})\nonumber \\
&&\times\, \left[\frac{eg^{(1)}_{N^{*}N\gamma}}{2m_N}\gamma_{\beta}
+\frac{eg^{(2)}_{N^{*}N\gamma}}{(2m_N)^2}p_{2\beta}\right]f_s(p_{s},m_{N(1520)}),\\[6pt]
M_{\rho}^{\mu\nu}&=& - \,e\frac{g_{RN\rho}g_{\pi\rho\gamma}}{2m_\pi m_N}\epsilon^{\alpha\mu\lambda\delta}p_{1\alpha}p_{3\lambda}S^{1}_{\delta\sigma}(p_{t})\nonumber \\
&&\times\, \gamma_{\beta}(p_t^{\nu}g^{\sigma\beta}-p_{t}^{\beta}g^{\sigma\nu})f_t(p_{t},m_{\rho}),\\[6pt]
M_{\omega}^{\mu\nu}&=& - \,e\frac{g_{RN\omega}g_{\pi\omega\gamma}}{2m_\pi m_N}\epsilon^{\alpha\mu\lambda\delta}p_{1\alpha}p_{3\lambda}S^{1}_{\delta\sigma}(p_{t})\nonumber \\
&&\times\, \gamma_{\beta}(p_t^{\nu}g^{\sigma\beta}-p_{t}^{\beta}g^{\sigma\nu})f_t(p_{t},m_{\omega}),\\[6pt]
M_{c}^{\mu\nu}&=&e\frac{g_{RN\pi}}{m_\pi}g^{\mu\nu}\gamma_{5}f_s(p_{s},m_{N}),
\end{eqnarray}
where $p_s=p_1+p_2$, $p_t=p_3-p_1$, and $p_u=p_2-p_3$.

Similarly, the standard amplitudes for the decay cascade $\gamma p \to \pi^{0} N(1710)1/2^{+} \to \pi^{0}\eta p$ can be expressed as
\begin{eqnarray}
M^{\mu}(L)=T M^{\mu}(R),
\end{eqnarray}
where the L and R in the parentheses are simplified for left and right, respectively; $T$ represents the $N(1710)1/2^{+} \to \eta p$ decaying part, which is given by
\begin{eqnarray}
T=-i\frac{g_{RN\eta}}{m_{R}+m_{N}} \gamma_{5}\slashed{p}_{4}S^{1/2}(p_{R}).
\end{eqnarray}

The $M^{\mu}(R)$ represent the remaining parts of the leg-amputated amplitudes $M^{\mu}(L)$, and they are expressed as
\begin{eqnarray}
M_{N(s)}^{\mu}&=& -\,i e \frac{g_{RN\pi}}{m_{N}-m_{R}}\gamma_{5}\slashed{p}_{3}S^{1/2}(p_{s}) \nonumber \\
&&\times \, \left(\gamma^{\mu}
+\frac{i\kappa_{N}}{2m_{N}}
\sigma^{\mu\beta}p_{1\beta}\right)f_{s}(p_s,m_N),   \\[6pt]
M_{N(u)}^{\mu}&=& e \frac{g_{NN\pi}g_{RN\gamma}}{{(2m_N)}^2}\sigma^{\mu\alpha}p_{1\alpha}S^{1/2}(p_{u})\nonumber \\
&&\times \,\gamma_{5}\slashed{p}_{3}f_{u}(p_u,m_N), \\[6pt]
M_{\Delta}^{\mu}&=& i\,\frac{g_{R\Delta\pi}}{m_\pi}p_{3}^{\alpha}S^{3/2}_{\alpha\delta}(p_{s})
(p_{1}^{\delta}g^{\mu\beta}-p_{1}^{\beta}g^{\mu\delta})\nonumber \\
&&\times \,\left[\frac{eg^{(1)}_{RN\gamma}}{2m_{N}}
\gamma_{\beta}\gamma_{5}+\frac{eg^{(2)}_{\Delta N\gamma}}{(2m_{N})^2}\gamma_{5}p_{2\beta} \right] f_{s}(p_{s},m_{\Delta}),  \\[6pt]
M_{N(1535)}^{\mu}&=& e\frac{g_{N^{*}R\pi}^{(1)}g_{N^{*}N\gamma}^{(1)}}{2m_{N}(m_{R}-m_{N})}\slashed{p}_{3}S^{1/2}(p_s)\nonumber \\
&&\times \, \gamma_{5}\sigma^{\mu\alpha}p_{1\alpha}f_s(p_{s},m_{N(1535)}), \\[6pt]
M_{\rho}^{\mu}&=& e\frac{g_{RN\rho}g_{\pi\rho\gamma}}{2m_\pi m_N}\left[\frac{-\gamma^{\beta}p_t^{2}}{m_R-m_N}-p_{t}^{\beta}\right]S^{1}_{\beta\delta}(p_{t})\nonumber \\
&&\times \, \epsilon^{\alpha\mu\lambda\delta}p_{1\alpha}p_{3\lambda}f_t(p_{t},m_{\rho}),\\[6pt]
M_{\omega}^{\mu}&=& e\frac{g_{RN\omega}g_{\pi\omega\gamma}}{2m_\pi m_N}\left[\frac{-\gamma^{\beta}p_t^{2}}{m_R-m_N}-p_{t}^{\beta}\right]S^{1}_{\beta\delta}(p_{t})\nonumber \\
&&\times \,\epsilon^{\alpha\mu\lambda\delta}p_{1\alpha}p_{3\lambda}f_t(p_{t},m_{\omega}),\\[6pt]
M_{c}^{\mu}&=&ie\frac{g_{RN\pi}}{m_N+m_R}\gamma_{5}\gamma^{\mu}f_s(p_{s},m_{N}).
\end{eqnarray}

The total amplitude $M$ is obtained by summing the individual amplitudes. It should be noted that due to the introduction of the form factors, the total amplitude $M$ is not gauge invariant in its current form. Specifically, for the $\gamma p \to \pi^0 N(1710)1/2^+ \to \pi^0 \eta p$ process, the $s$-channel $N$ pole contribution and the $u$-channel $N$ pole contribution breaks gauge invariance. Similarly, for the $\gamma p \to \pi^0 N(1700)3/2^- \to \pi^0 \eta p$ process, the $s$-channel $N$ pole contribution also breaks gauge invariance. In our previous studies \cite{ff3,ac2018}, we employed a generalized gauge-invariant term following the approach outlined in Refs.~\cite{Haberzettl:1997,Haberzettl:2006} to ensure gauge invariance of the amplitude. However, given the multiple approximations adopted in the present work, we focus on maintaining the simplicity of the formalism. As a result, instead of incorporating the generalized gauge-invariant term, a general contact term has been introduced. It should be noted that, unlike the generalized gauge-invariant term, the general contact term does not preserve the gauge invariance of the amplitude. This limitation is deemed acceptable within the scope of the present study, given the emphasis on simplicity and practical approximations.

The theoretical results for differential cross-sections, total cross-sections, and invariant mass distributions can be calculated using the formula

\begin{eqnarray}
d\sigma &=& \frac{1}{4(2\pi)^5(p_{1}\cdot p_{2})} \sum_{\lambda_{\gamma}\lambda_{p}\lambda_{p'}}|M|^2\frac{d^3p_3}{2E_{\pi}}\frac{d^3p_4}{2E_{\eta}}
\frac{d^3p_5}{2E_{p'}}\nonumber \\[6pt]
&&\times \delta^{4}(p_1+p_2-p_3-p_4-p_5),
\end{eqnarray}
where $\lambda_\gamma$, $\lambda_{p}$ and $\lambda_{p'}$ are the helicities of the photon, incoming proton and outgoing proton, respectively; $p_1$, $p_2$, $p_3$, $p_4$, $p_5$ are the four-momenta of the photon, incoming proton, $\pi$, $\eta$, and outgoing proton, respectively.

\section{Results and discussion}   \label{Sec:results}

\begin{figure*}[tbp]
\includegraphics[width=0.8\textwidth]{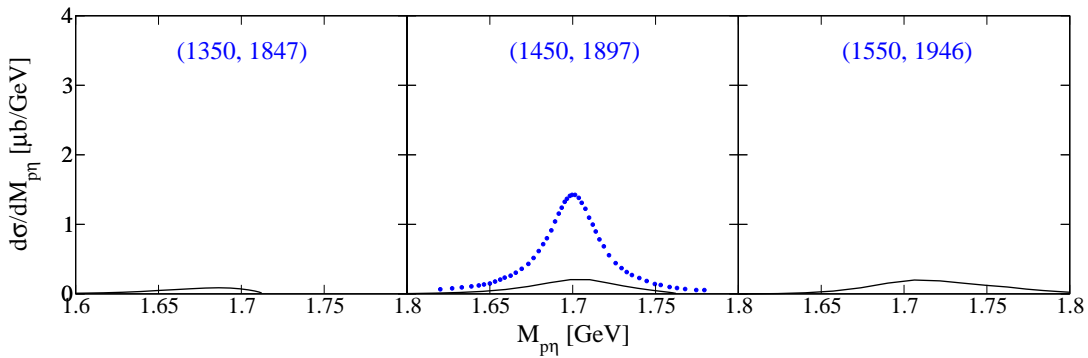}
\caption{The invariant mass distributions of decay cascade $\gamma p \to \pi^{0} N(1700)3/2^{-} \to \pi^{0}\eta p$ as a function of the invariant mass of $\eta p$ (black solid lines). The blue scattered symbols denote the individual contribution of the narrow structure stripped by the CBELSA/TAPS Collaboration \cite{xin:2021}. The numbers in parentheses denote the centroid value of the photon laboratory incident energy (left number) and the corresponding total center-of-mass energy of the system (right number), in MeV.}
\label{fig:inva1}
\end{figure*}

\begin{figure*}[tbp]
\includegraphics[width=0.8\textwidth]{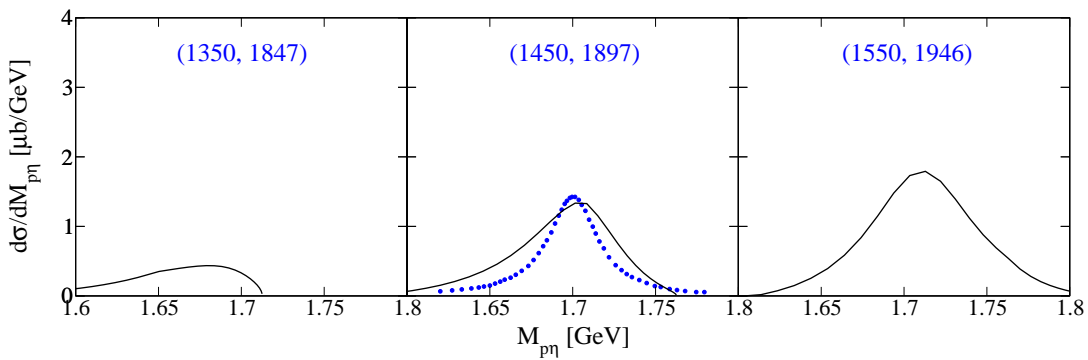}
\caption{The invariant mass distributions of decay cascade $\gamma p \to \pi^{0} N(1710)1/2^{+} \to \pi^{0}\eta p$ as a function of the invariant mass of $\eta p$ (black solid lines). The notations are the same as in Fig.~\ref{fig:inva1}.}
\label{fig:inva2}
\end{figure*}

\begin{figure}[tbp]
\includegraphics[width=\columnwidth]{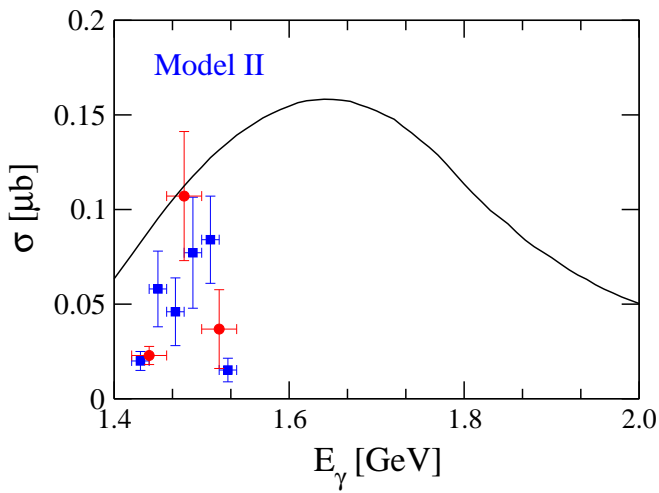}
\caption{The cross sections of the decay cascade $\gamma p \to \pi^{0} N(1710)1/2^{+} \to \pi^{0} \eta p$ as a function of the incident photon energy (black solid line). The blue and red scattered symbols correspond to the individual contributions of the narrow structure extracted by the CBELSA/TAPS Collaboration \cite{xin:2021}.}
\label{fig:total2}
\end{figure}

Following the discovery of the narrow structure in the $\eta N$ invariant mass distributions at $W \sim 1.68$ GeV for the $\gamma N \to \pi \eta N$ reaction at the GRAAL facility \cite{jiu:2017}, the CBELSA/TAPS Collaboration remeasured the $\gamma p \to \pi^{0} \eta p$ reaction and observed the narrow structure at a different energy $W \sim 1.7$ GeV \cite{xin:2021}. Two contrasting explanations have been proposed for this narrow structure. In the first interpretation \cite{jiu:2017}, it is suggested that the narrow structure results from a new isospin $1/2$ nucleon resonance, denoted as $N(1685)$. In contrast, the CBELSA/TAPS Collaboration, in their 2021 study \cite{xin:2021}, argued that the narrow structure is more likely caused by the triangular singularity in the $\gamma p \to \pi^{0}\eta p$ reaction rather than the decay cascade via an intermediate resonance.

To clarify the nature of this narrow structure, it is essential to distinguish between the contribution of the narrow structure and the dominant background originating from the $\gamma p \to \Delta(1232)\eta \to p\pi^0\eta$ mechanism, as identified by the CBELSA/TAPS Collaboration \cite{xin:2021}. In their analysis, the narrow structure signal (represented by the red solid curve in Fig.~7 of Ref.~\cite{xin:2021}) was extracted from the background dominated by the $\Delta(1232)$ resonance (depicted by the blue dotted curve in Fig.~7 of Ref.~\cite{xin:2021}). Our present work focuses exclusively on reproducing the extracted narrow structure contribution, rather than the entire $M_{p\eta}$ invariant mass spectrum. By isolating the narrow structure, we aim to provide a quantitative assessment of whether the observed structure could result from the decay cascade via an intermediate nucleon resonance. Additionally, we explore the possible roles of decay cascades involving other intermediate resonances, beyond the dominant $\Delta(1232)$.

At the energy where the narrow structure was detected, approximately $W \sim 1.7$ GeV, the PDG review \cite{PDG:2022} lists two potential nucleon resonances: the evaluated three-star $N(1700)3/2^-$ and the four-star $N(1710)1/2^+$. Both resonances have decay branching ratios to the $\eta p$ channel. Specifically, according to the PDG review, the decay branching ratio to $\eta p$ is estimated to be $1-2\%$ for the $N(1700)3/2^-$ and $10-50\%$ for the $N(1710)1/2^+$. For this study, we take the medians of these estimated ranges. The masses of the nucleon resonances $N(1700)3/2^-$ and $N(1710)1/2^+$ are taken as $1700$ MeV and $1710$ MeV, respectively. Our model intentionally excludes contributions from the $\gamma p \to \Delta(1232)\eta \to p\pi^0\eta$ mechanism, as these contribute primarily to the background rather than the narrow structure of interest.

Regarding the width values, the PDG provides estimated ranges of $100-300$ MeV for the $N(1700)3/2^-$ and $80-200$ MeV for the $N(1710)1/2^+$. However, in our present work, we opt for a width of $80$ MeV for both nucleon resonances. This choice is motivated by the observed behavior of the narrow structure in the CBELSA/TAPS Collaboration's data \cite{xin:2021}. As the incident energy increases from $1420$ MeV to $1540$ MeV, the width of the observed structure grows from $35$ MeV to about $50$ MeV and then decreases. The chosen value of $80$ MeV aims to strike a balance between the advocated width values by the PDG and the experimental observations, taking into consideration the potential interference effects between the individual contribution of the narrow structure and its background.

To investigate whether the narrow structure in the $\eta p$ invariant mass distributions can be attributed to the decay cascade via intermediate nucleon resonance decaying into $\eta p$ final states, we employ an effective Lagrangian method at the tree-level approximation. Our study considers all relevant terms based on decay information from the PDG review \cite{PDG:2022}. In addition to the background terms, including $t$-channel $\rho$ and $\omega$ exchanges, $s$-channel $N$ and $\Delta$ pole contributions, $u$-channel $N$ exchange, and the contact term, we include the pole contributions of $s$-channel nucleon resonances coupled to the $\pi Res.$ channel. According to the PDG review, for the case of $\gamma p \to \pi^{0} N(1700)3/2^{-} \to \pi^{0}\eta p$, we consider the nucleon resonances $N(1440)1/2^{+}$ and $N(1520)3/2^{-}$ in the $s$ channel. For the case of $\gamma p \to \pi^{0} N(1710)1/2^{+} \to \pi^{0}\eta p$, we consider the nucleon resonance $N(1535)1/2^-$ in the $s$ channel.

\begin{table}[tb]
\caption{\label{Table:para} Model parameters. The same value is taken for the cutoff mass of the $t$-channel $\rho$ and $\omega$ exchanges, and the same value is taken for the cutoff mass of the other terms. }
\begin{tabular*}{\columnwidth}{@{\extracolsep\fill}lrrrr}
\hline\hline
                                &  Model I  &  Model II  \\
\hline

$\Lambda_{\rho,\omega}$ [MeV]   &$2500$     & $1550$     \\
$\Lambda_{others}$ [MeV]        &$2500$     & $2500$     \\
\hline\hline
\end{tabular*}
\end{table}

As detailed in Sec.~\ref{Sec:formalism}, all coupling parameters in our study are fixed using decay branching ratios or helicity amplitudes referenced from the PDG review \cite{PDG:2022}, providing a rigorous foundation for constructing the theoretical framework. The cutoff masses are treated as fitting parameters and are summarized in Table~\ref{Table:para}. Table~\ref{Table:para} reveals that we use the same cutoff mass value for the $t$-channel $\rho$ and $\omega$ exchanges, and another common cutoff mass value for the remaining terms.

The theoretical results for the invariant mass distributions of the $\gamma p \to \pi^{0}N(1700)3/2^{-} \to \pi^{0}\eta p$ decay cascade reaction, corresponding to the parameters of Model I listed in the second column of Table~\ref{Table:para}, are presented in Fig.~\ref{fig:inva1}. The black solid lines represent the theoretical predictions for the narrow structure contribution, while the blue scattered symbols correspond to the experimentally extracted narrow structure signal, as provided by the CBELSA/TAPS Collaboration \cite{xin:2021}. It is important to note that the experimental data have been processed with a cut of $M_{p\pi^{0}}<1190$ MeV to suppress the dominant background from the decay cascade via the $\Delta(1232)$ resonance, which dominates in the $\gamma p \to \pi^{0}\eta p$ reaction. Our theoretical calculations also incorporate this cut, ensuring consistency with the experimental conditions and focusing on reproducing the narrow structure contribution.

As depicted in Fig.~\ref{fig:inva1}, the theoretical results fall short of reaching the order of magnitude of the experimental data. In other words, the decay cascade via the intermediate nucleon resonance $N(1700)3/2^{-}$ cannot account for the narrow structure observed in the $\eta p$ invariant mass distributions. The primary reason for this discrepancy is the weak coupling of the nucleon resonance $N(1700)3/2^{-}$ to $\eta p$ final states, with a decay branching ratio of $1-2\%$ as indicated by the PDG review \cite{PDG:2022}. This results in the theoretical results having an order of magnitude lower than that of the data.

The theoretical results for the invariant mass distributions of the $\gamma p \to \pi^{0}N(1710)1/2^{+}\to \pi^{0}\eta p$ decay cascade, corresponding to the parameters of Model II as listed in the third column of Table~\ref{Table:para}, are presented in Fig.~\ref{fig:inva2}. Figure~\ref{fig:inva2} demonstrates that the decay cascade via the intermediate nucleon resonance $N(1710)1/2^{+}$ provides a qualitative description of the stripped experimental curve. However, there is room for improvement in the description. The fixed width of the nucleon resonance $N(1710)1/2^{+}$ is $80$ MeV, which surpasses that of the observed structure represented as $35$ MeV \cite{xin:2021}. The interference effect between the contribution of the narrow structure and its background terms introduces imprecision to the stripped contribution of the narrow structure. Thus, the current descriptive quality of Model II for the invariant mass distributions, as illustrated in Fig.~\ref{fig:inva2}, is considered acceptable.

Nevertheless, as shown in Fig.~\ref{fig:total2}, with increasing incident energy, the cross section of the decay cascade $\gamma p \to \pi^{0}N(1710)1/2^{+}\to \pi^{0}\eta p$ deviate significantly from the CBELSA/TAPS data of the narrow structure \cite{xin:2021}. The observed data exhibit a much smaller cross-section width, and this substantial deviation cannot be solely attributed to the interference effects between the narrow structure and its background terms, or the fixed width of $80$ MeV. Consequently, we conclude that, regardless of whether the intermediate nucleon resonance is $N(1700)3/2^-$ or $N(1710)1/2^+$, the decay cascade via the intermediate nucleon resonance cannot account for the narrow structure observed in the $\eta p$ invariant mass distributions.

\section{Summary and conclusion}  \label{Sec:summary}

The CBELSA/TAPS Collaboration detected a narrow structure in the $\eta p$ invariant mass distributions in the $\gamma p \to \pi^{0} \eta p$ reaction at an incident energy of $W \sim 1.7$ GeV \cite{xin:2021}. Two distinct explanations have been proposed: the structure is attributed either to a new isospin-$1/2$ nucleon resonance $N(1685)$ \cite{jiu:2017} or to the triangular singularity mechanism \cite{xin:2021}. Motivated by these interpretations, we performed a quantitative analysis to assess whether the observed narrow structure could result from the decay cascade via an intermediate nucleon resonance in the $\gamma p \to \pi^{0}Res. \to \pi^{0} \eta p$ process.

Our study considers the effective Lagrangian method at tree level, incorporating contributions from nucleon resonances $N(1700)3/2^-$ and $N(1710)1/2^+$, which are listed in the PDG review \cite{PDG:2022} as potential candidates near the energy of interest. The theoretical model includes contributions from $s$-channel pole diagrams for intermediate resonances as well as $t$-channel $\rho$ and $\omega$ exchanges, $s$- and $u$-channel nucleon and $\Delta$ pole contributions, and a contact term. Specific intermediate resonance decay processes, such as $N(1440)1/2^{+}$ and $N(1520)3/2^{-}$ for $\gamma p \to \pi^{0}N(1700)3/2^{-} \to \pi^{0}\eta p$, and $N(1535)1/2^-$ for $\gamma p \to \pi^{0}N(1710)1/2^{+} \to \pi^{0}\eta p$, are explicitly included.

Theoretical results show that the contribution of the decay cascade $\gamma p \to \pi^{0}N(1700)3/2^{-}\to \pi^{0}\eta p$ is orders of magnitude smaller than the experimentally extracted narrow structure signal \cite{xin:2021}, making $N(1700)3/2^-$ an unlikely candidate. For $N(1710)1/2^+$, although a qualitative description of the invariant mass distribution of the narrow structure is achieved, the theoretical cross-section width is significantly larger than the experimental results. Thus, $N(1710)1/2^+$ cannot adequately explain the observed narrow structure either. These findings indicate that the decay cascade via intermediate nucleon resonances, whether $N(1700)3/2^-$ or $N(1710)1/2^+$, cannot account for the narrow structure observed in the $\eta p$ invariant mass distributions of the $\gamma p \to \pi^{0}\eta p$ reaction.

In conclusion, the observed narrow structure in the $\eta p$ invariant mass distributions is unlikely to arise from decay cascades involving intermediate resonances. Among the candidates investigated, $N(1710)1/2^+$ contributes more than $N(1700)3/2^-$ but remains insufficient to explain the narrow structure. This result supports the understanding that the $\gamma p \to \pi^{0}\eta p$ reaction is predominantly governed by decay cascades via the $\Delta(1232)$ resonance, as suggested in previous studies.

\begin{acknowledgments}
The authors thank Fei Huang for useful discussions. This work is partially supported by the National Natural Science Foundation of China under Grants No.~12305097, and No.~12305137, the Fundamental Research Funds for the Central Universities (Grant No.~23CX06037A), the Shandong Provincial Natural Science Foundation, China (Grant No.~ZR2024QA096), and the Taishan Scholar Young Talent Program (Grant No.~tsqn202408091).
\end{acknowledgments}

\end{document}